\documentclass[fleqn,usenatbib]{mnras}

\usepackage{newtxtext,newtxmath}

\usepackage[T1]{fontenc}

\DeclareRobustCommand{\VAN}[3]{#2}
\let\VANthebibliography\thebibliography
\def\thebibliography{\DeclareRobustCommand{\VAN}[3]{##3}\VANthebibliography}

\usepackage{graphicx}	
\usepackage{amsmath}	

\title[Enhancement of Li in RC Stars]{Enhancement of Lithium in Red Clump Stars by the Additional Energy Loss Induced by New Physics}

\author[K. Mori et al.]{
Kanji Mori,$^{1,2}$\thanks{E-mail: kanji.mori@grad.nao.ac.jp}
Motohiko Kusakabe,$^{3}$
A. Baha Balantekin,$^{4,2}$
Toshitaka Kajino,$^{2,3,1}$
and Michael A. Famiano$^{5,2}$
\\
$^{1}$Graduate School of Science, The University of Tokyo, 7-3-1 Hongo, Bunkyo-ku, Tokyo, 113-0033 Japan\\
$^{2}$Division of Science, National Astronomical Observatory of Japan, 2-21-1 Osawa, Mitaka, Tokyo 181-8588, Japan\\
$^{3}$School of Physics, Beihang University, 37 Xueyuan Road, Haidian-qu, Beijing 100083, China\\
$^4$Department of Physics, University of Wisconsin-Madison, Madison, Wisconsin 53706 USA\\
$^5$Department of Physics, Western Michigan University, Kalamazoo, Michigan 49008 USA\\
}

\date{Accepted XXX. Received YYY; in original form ZZZ}

\pubyear{2020}

\begin{document}
\label{firstpage}
\pagerange{\pageref{firstpage}--\pageref{lastpage}}
\maketitle

\begin{abstract}
Since $^7$Li is easily destroyed in low temperatures, the surface lithium abundance decreases as stars evolve. This is supported by the lithium depletion observed in the atmosphere of most red giants. However, recent studies show that almost all of red clump stars have high lithium abundances $A(\mathrm{Li})>-0.9$, which are not predicted by the standard theory of the low-mass stellar evolution. In order to reconcile the discrepancy between the observations and the model, we consider additional energy loss channels which may come from physics beyond the Standard Model. $A(\mathrm{Li})$ slightly increases near the tip of the red giant branch even in the standard model with thermohaline mixing because of the $^7$Be production by the Cameron-Fowler mechanism, but the resultant $^7$Li abundance is much lower than the observed values. We find that the production of $^7$Be becomes more active if there are additional energy loss channels, because themohaline mixing becomes more efficient and  a heavier helium core is formed. 
\end{abstract}




\section{Introduction}

Since $^7$Li is a fragile nucleus which is easily destroyed by the proton capture reaction, its surface abundance reflects detailed stellar structure. In low-mass giants, stellar models predict surface lithium depletion \citep{1967ApJ...147..624I}. However, spectroscopic surveys have shown that $\sim1\%$ of giant stars have the lithium abundance as high as $A(\mathrm{Li})=\log(\mathrm{Li/H})+12>1.5$ \citep[e.g.][]{2016MNRAS.461.3336C,2018NatAs...2..790Y,2018A&A...617A...4S,2019MNRAS.484.2000D}. This is a long-standing problem in our understanding of low-mass stars \citep{1982ApJ...255..577W,1989ApJS...71..293B}.

Stars in the red giant (RG) branch and the red clump (RC) have the similar luminosity and the effective temperature, so the boundary between them is ambiguous in the Hertzsprung-Russell diagram. Some authors have suggested that a  part of the lithium-rich giants are RC stars \citep{2014ApJ...784L..16S,2014A&A...564L...6M}. Recent works  \citep{2019MNRAS.482.3822S,2019ApJ...878L..21S,2020NatAs.tmp..139K}  distinguished RC stars from RGs in data of spectroscopic surveys with the help of asteroseismological data \citep{2011Natur.471..608B,2016A&A...588A..87V}. They concluded that all of RC stars have the  lithium abundances of $A(\mathrm{Li})>-0.9$, which are higher than the predicted values by stellar models. This implies that a ubiquitous process produces $^7$Li during or before central helium burning.

The mechanism of this lithium enhancement is under debate. Some authors suggest engulfment of substellar objects which keep high lithium abundances \citep[e.g.][]{1999MNRAS.308.1133S,2012A&A...538A..36L,2016ApJ...833L..24A}. Others discuss in situ production by the Cameron-Fowler (CF) mechanism \citep{1971ApJ...164..111C}. In the hydrogen burning shell, $^7$Be is produced via the $^3$He$(\alpha,\;\gamma)^7$Be reaction. The produced $^7$Be is conveyed to the stellar surface and decays to $^7$Li by the electron capture. In the standard model, the CF mechanism is insufficient to reproduce the abundance in lithium-rich giants. However, \citet{2016MNRAS.461.3336C} point out that extra mixing induced by the tidal interaction with a binary companion can drive the lithium production. Also,  a recent study \citep{2020arXiv200901248S} focused on mixing induced by the helium flash and claimed that it can naturally solve the lithium problem in RC stars proposed by \citet{2020NatAs.tmp..139K}.

In order to explain the ubiquitous enhancement of lithium in RC stars, we introduce the additional energy losses induced by physics beyond the Standard Model. In this {paper}, as an example we focus on the neutrino magnetic moment (NMM), which is denoted as $\mu_\nu$. The existence of neutrino masses was established by the detection of the neutrino oscillations, starting with atmospheric neutrinos  \citep{1998PhRvL..81.1562F} and later by solar and reactor neutrinos. In the Standard Model of particle physics  massive neutrinos have magnetic moments which are too small to be detected by present and near-term future experiments  \citep{1980PhRvL..45..963F,1982NuPhB.206..359S,2015RvMP...87..531G,2018ARNPS..68..313B}. The current best experimental limit $\mu_\nu<2.9\times10^{-11}\mu_\mathrm{B}$, where $\mu_\mathrm{B}$ is the Bohr magneton, comes from the GEMMA experiment \citep{2013PPNL...10..139B}, which measured the scattering cross sections of target electrons and reactor anti-electron neutrinos. 

The NMM induces the additional energy loss and affects various stages of
stellar evolution for wider mass-range of stars.  The effect on the
evolution of intermediate-mass stars has recently been studied in detail 
\citep{2020arXiv200808393M} and it was found that in the presence of a sufficiently large NMM the duration of blue giants is
shorter and the blue loops are eliminated.  Stellar plasma of low-mass
stars also is effected by NMM and the helium flash delays \citep{1994ApJ...425..222H}. As a result, a heavier inert helium core is formed and the luminosity of the tip of the RG branch (TRGB) increases \citep[e.g.][]{1996slfp.book.....R}. This enables one to use low-mass stars in globular clusters to give tighter constraints of $\mu_\nu<1.5\times10^{-12}\mu_\mathrm{B}$ \citep{2020arXiv200703694C} and $\mu_\nu<2.2\times10^{-12}\mu_\mathrm{B}$ \citep{2015APh....70....1A}. Also, the delayed helium flash may result in activation of the CF mechanism induced by thermohaline mixing \citep{1999ApJ...510..217S,2015MNRAS.446.2673L}. 
In this paper we show that NMM needed to enhance $A(\mathrm{Li})$ in RC stars and  reduce the discrepancy between the observations and the theory is slightly higher than current astrophysical limits, so other  channels may be in the play. 

Section 2 describes the stellar models and the treatment of the NMM. Section 3 shows the results of our  calculations and compares them with the observational data. In Section 4, we summarize our results and discuss the future perspective.

\section{Methods}

We use Modules for Experiments in Stellar Astrophysics \citep[\texttt{MESA};][]{MESA1,MESA2,MESA3,MESA4,MESA5} version 10398 to construct one-dimensional low-mass stellar models. \texttt{MESA} adopts the equation of state of \citet{Rogers2002} and \citet{Timmes2000} and the opacity of \cite{Iglesias1996,Iglesias1993} and \cite{Ferguson2005}. The electron conductivity is from \citep{2007ApJ...661.1094C}.  We adopt nuclear reaction rates compiled by  NACRE \citep{1999NuPhA.656....3A} and \citet{1988ADNDT..40..283C}. If a reaction rate appears in both, the one tabulated in NACRE is adopted. {In particular}, the rates for the triple-$\alpha$ reaction and $^{14}$N$(p,\;\gamma)^{15}$O are from NACRE.  The adopted nuclear reaction network is \texttt{pp\_and\_cno\_extras.net}, which includes $^{1,2}$H, $^{3,4}$He, $^7$Li, $^7$Be, $^8$B, $^{12}$C, $^{14}$N, $^{14,16}$O, $^{19}$F, $^{18,19,20}$Ne, and $^{22,24}$Mg.  Treatment of  electron screening is based on for the strong regime and \citet{1973ApJ...181..457G} for the weak regime.  The mass loss formula in \citet{1975MSRSL...8..369R} with the scaling factor $\eta=0.3$ is adopted.

\texttt{MESA} makes use of the mixing length theory \citep{1968pss..book.....C} to calculate the convective luminosity from the temperature gradient. We adopt the Ledoux criterion \citep{1947ApJ...105..305L} to calculate the convective instability.  Our model considers thermohaline mixing as well because it affects the lithium abundance \citep{1999ApJ...510..217S,2015MNRAS.446.2673L}. Thermohaline mixing is treated as a diffusive process with the diffusion coefficient \citep{MESA2}
\begin{equation}
    D_\mathrm{thm}=\alpha_\mathrm{thm}\frac{3K}{2\rho C_\mathrm{P}}\frac{B}{\nabla_\mathrm{T}-\nabla_\mathrm{ad}},
\end{equation}
where $\alpha_\mathrm{thm}$ is a free parameter, $\rho$ is the density, $C_\mathrm{P}$ is the specific heat, $B$ is the Ledoux term \citep{1989nos..book.....U}, $\nabla_\mathrm{T}$ is the actual temperature gradient, and $\nabla_\mathrm{ad}$ is the adiabatic temperature gradient. $K$ is the thermal conductivity written as
\begin{equation}
        K=\frac{4acT^3}{3\kappa\rho},\label{K}
\end{equation}
where $a$ is the radiation density constant, $c$ is the speed of light, $T$ is the temperature, and $\kappa$ is the opacity.


The parameters in our models follow those in \citet{2020NatAs.tmp..139K}. The initial mass is fixed to $1M_{\odot}$ and the initial metallicity is fixed to be solar: $Z=0.0148$ \citep{2019arXiv191200844L}. However, the initial lithium abundance in the pre-main sequence is set to $A(\mathrm{Li})=2.8$. Although there is no observational evidence, this value is used in previous works \citep{2020NatAs.tmp..139K,2020arXiv200901248S}. We follow their prescription to compare our result with theirs.  The mixing length is $\alpha=1.6$ and the thermohaline coefficient is $\alpha_\mathrm{thm}=50$, 100 and 1000, because $A(\mathrm{Li})$ after the RG branch bump is sensitive to thermohaline mixing.

We consider the plasmon decay and neutrino pair production as the additional energy loss induced by the NMM. The energy loss rate due to the plasmon decay is given by \citep{1994ApJ...425..222H,2009ApJ...696..608H}
\begin{equation}
\epsilon^\mu_\mathrm{plas}=0.318\left(\frac{\omega_\mathrm{pl}}{10\;\mathrm{keV}}\right)^{-2}\left(\frac{\mu_\nu}{10^{-12}\mu_\mathrm{B}}\right)^2\epsilon_\mathrm{plas},\label{plas}
\end{equation}
where $\epsilon_\mathrm{plas}$ is the standard plasmon decay  rate \citep{1996ApJS..102..411I} and $\omega_\mathrm{pl}$ is the plasma frequency \citep{1996slfp.book.....R}
\begin{equation}
    \omega_\mathrm{pl}=28.7\;\mathrm{eV}\frac{(Y_\mathrm{e}\rho)^\frac{1}{2}}{(1+(1.019\times10^{-6}Y_\mathrm{e}\rho)^\frac{2}{3})^\frac{1}{4}},
\end{equation}
where $Y_\mathrm{e}$ is the electron mole fraction and $\rho$ is the density in units of g cm$^{-3}$. The energy loss rate due to the pair production is given by \citep{2009ApJ...696..608H}
\begin{equation}
    \epsilon^\mu_\mathrm{pair}=1.6\times10^{11}\;\mathrm{erg\;g^{-1}\;s^{-1}}\left(\frac{\mu_\nu}{10^{-10}\mu_\mathrm{B}}\right)^2\frac{e^{-\frac{118.5}{T_8}}}{\rho_4},\label{pair}
\end{equation}
where $T_8$ is the temperature in units of 0.1 GK and $\rho_4=\rho/(10^4\;\mathrm{g\; cm^{-3}})$.
\begin{figure}
\centering
\includegraphics[width=8.5cm]{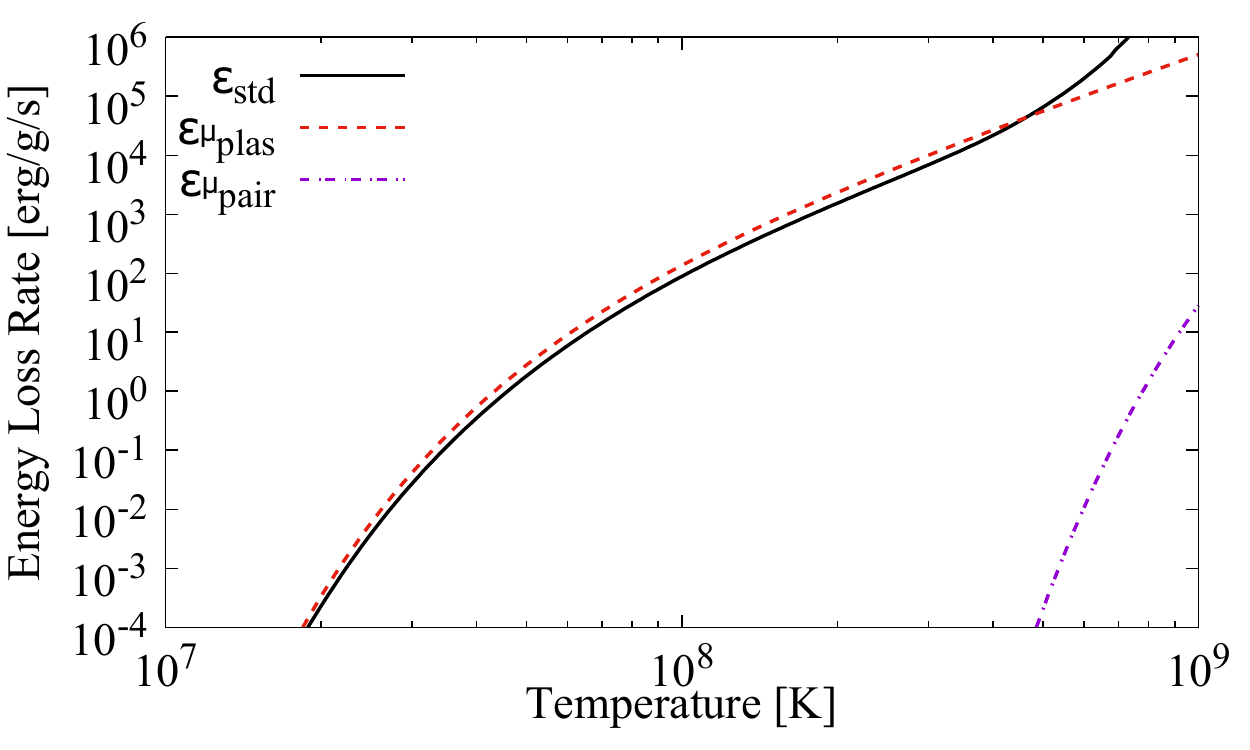}
\caption{The energy loss rates induced by the neutrino emission with $\mu_\nu=5\times10^{-12}\mu_\mathrm{B}$. The density of $\rho=10^6$ g cm$^{-3}$ is assumed. The black line shows the total standard rate and the other lines show the rates induced} by the NMM. The total energy loss rate is given by $\epsilon=\epsilon_\mathrm{std}+\epsilon^\mu_\mathrm{plas}+\epsilon^\mu_\mathrm{pair}$. \label{energy}
\end{figure}

Fig. \ref{energy} shows the energy loss rates with $\mu_\nu=5\times10^{-12}\mu_\mathrm{B}$ at $\rho=10^6$ g cm$^{-3}$, which is the typical central density at the helium flash. At the temperature of $\sim10^8$ K, the additional energy loss rate is comparable with the standard rate. Also, it is seen that the pair production is negligible in the temperature range of interest.

\section{Result}
\subsection{Evolution of the Fiducial Models}
\label{result1}
\begin{figure*}
\centering
\includegraphics[width=12cm]{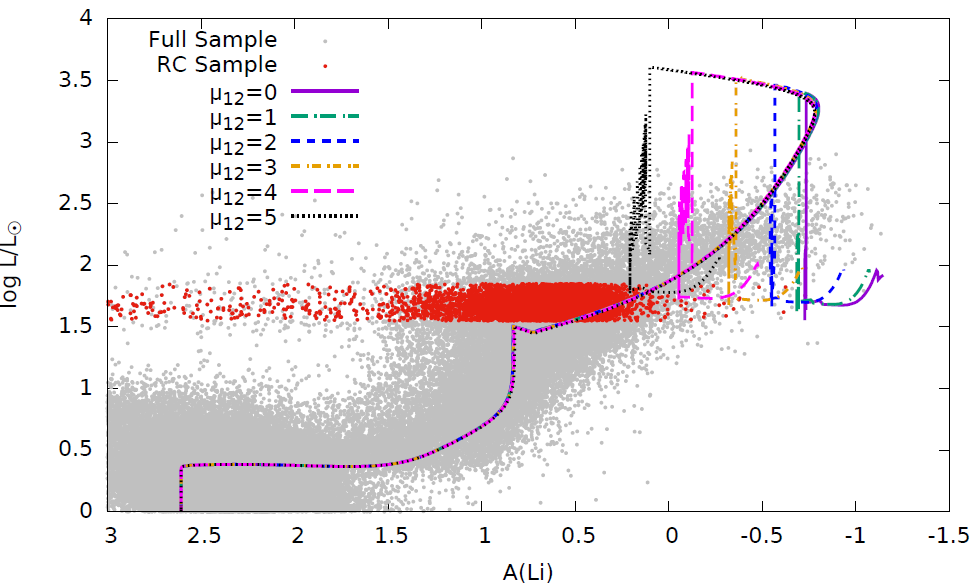}
\includegraphics[width=12cm]{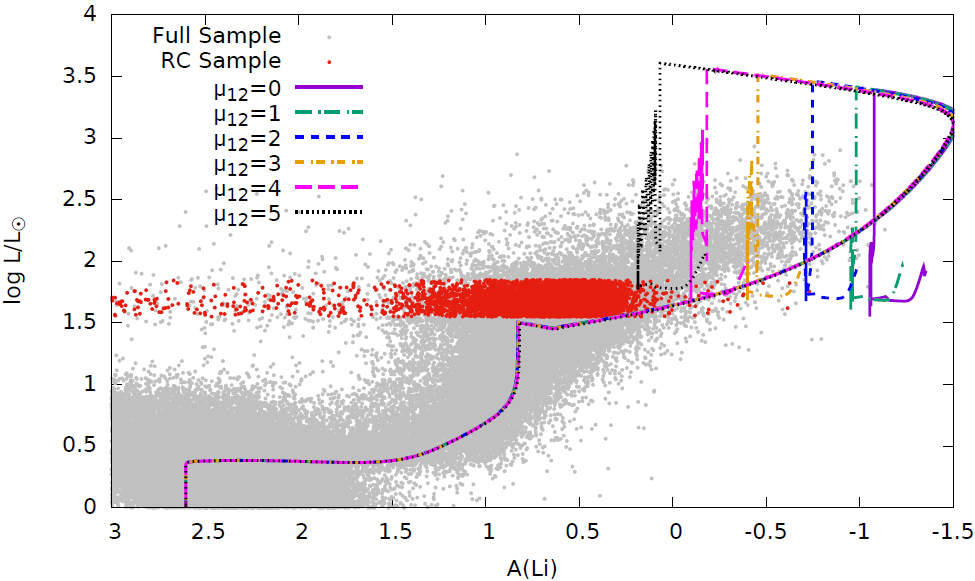}
\includegraphics[width=12cm]{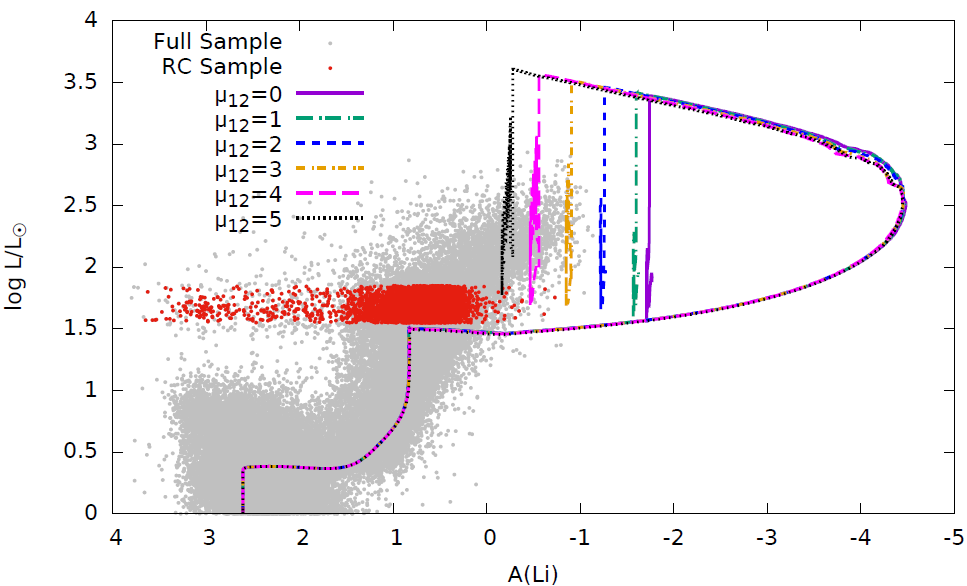}
\caption{The lines show the evolution of our models with $\mu_{12}=0-5$ in the $L-A$(Li) plane. The upper panel adopts $\alpha_\mathrm{thm}=50$, the middle panel adopts $\alpha_\mathrm{thm}=100$, and the lower panel adopts $\alpha_\mathrm{thm}=1000$. The grey dots are  GALAH DR2 samples \citep{2018MNRAS.478.4513B} with reliable lithium abundances and the red dots are RC samples selected by \citet{2020NatAs.tmp..139K}. }
\label{fig:lumi}
\end{figure*}

In this Section, we describe the fiducial model with $\mu_\nu=0$. Fig. \ref{fig:lumi} shows the evolution of the stellar models in the $L-A(\mathrm{Li})$ plane, where $L$ is the luminosity. The solid lines show the evolution of the fiducial model. The upper panel adopts $\alpha_\mathrm{thm}=50$, the middle panel adopts $\alpha_\mathrm{thm}=100$, and the lower panel adopts $\alpha_\mathrm{thm}=1000$.

The evolution starts from a low luminosity (the lower-left side of Fig. \ref{fig:lumi}). The lithium abundance $A(\mathrm{Li})$ stays constant during the main sequence. When the star reaches the main-sequence turnoff at $\log(L/L_\odot)=0.4$, $A(\mathrm{Li})$ starts to decrease from 2.6 to 0.8. This is because surface lithium is conveyed to the stellar interior due to the first dredge-up \citep{1967ApJ...147..624I}, and is destroyed by the proton capture. The lithium depletion becomes slower as the star evolves, but $A(\mathrm{Li})$ starts to decrease again when the star reaches the RG branch bump at $\log(L/L_\odot)=1.5$. At this point, the star develops thermohaline mixing between the convective envelope and the hydrogen burning shell \citep{2007A&A...467L..15C,2015MNRAS.446.2673L}. This happens because the mean molecular weight is inverted by $^3$He($^3$He, $2p)^4$He. Because of thermohaline mixing, lithium in the envelope is conveyed to the inner hot region and destroyed. One can see that $A$(Li) after the RG branch bump is smaller when a larger $\alpha_\mathrm{thm}$ is adopted. {In particular}, when $\alpha_\mathrm{thm}=1000$, lithium is depleted so much that the model cannot reproduce the observed samples. \citet{2015MNRAS.446.2673L} calculated $A$(Li) in RGs with $C_t=10^2$, $10^3$, and $10^4$, where $C_t=3\alpha_\mathrm{thm}/2$. {Although the value of A(Li) varies by changing a stellar evolution code adopted}, they also reported that a larger $\alpha_\mathrm{thm}$ leads to a larger variation of $A$(Li) during the evolution along the RG branch.

The decrease of $A(\mathrm{Li})$ stops when $\log(L/L_\odot)=3.2$ and it starts increasing. This is because thermohaline mixing becomes more efficient as the star expands \citep{2015MNRAS.446.2673L}. The efficient mixing helps the CF mechanism work and hence increases $A(\mathrm{Li})$. After the TRGB, the core becomes non-degenerate because of the helium flash and $L$ decreases suddenly.  As a result, core helium burning begins and a RC star is formed.

In order to understand the evolution of $A$(Li), we introduce three timescales. We denote the timescale for the electron capture of $^7$Be as $t_\mathrm{prod}$, the timescale for $^7$Li$(p,\;\alpha)^4$He as $t_\mathrm{dest}$, and the timescale on which the material at a given radius is conveyed to the bottom of the convective envelope by thermohaline mixing as $t_\mathrm{mix}$.
\begin{figure}
\centering
\includegraphics[width=8.5cm]{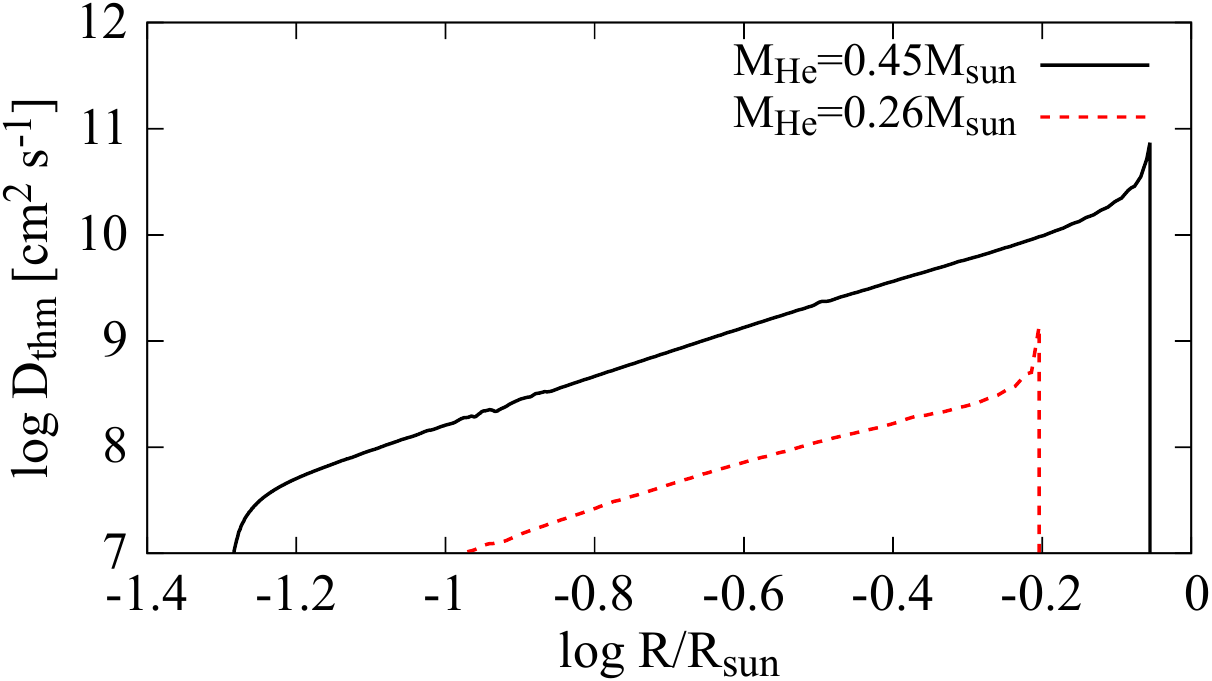}
\includegraphics[width=8.5cm]{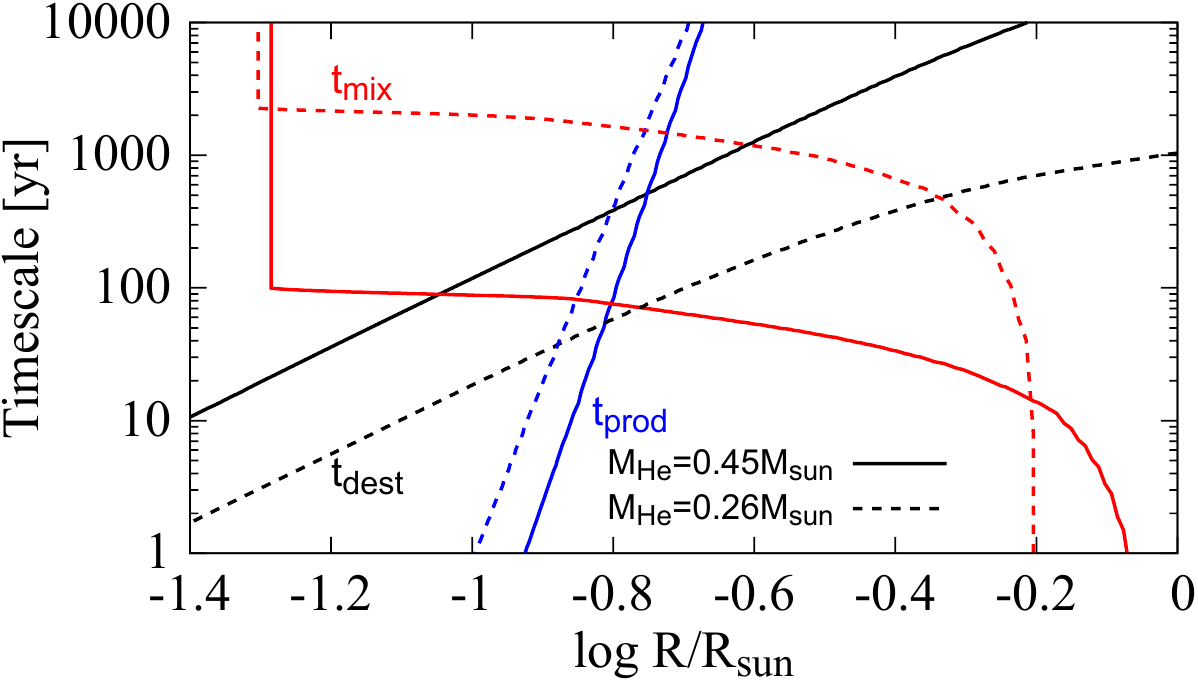}
\includegraphics[width=8.5cm]{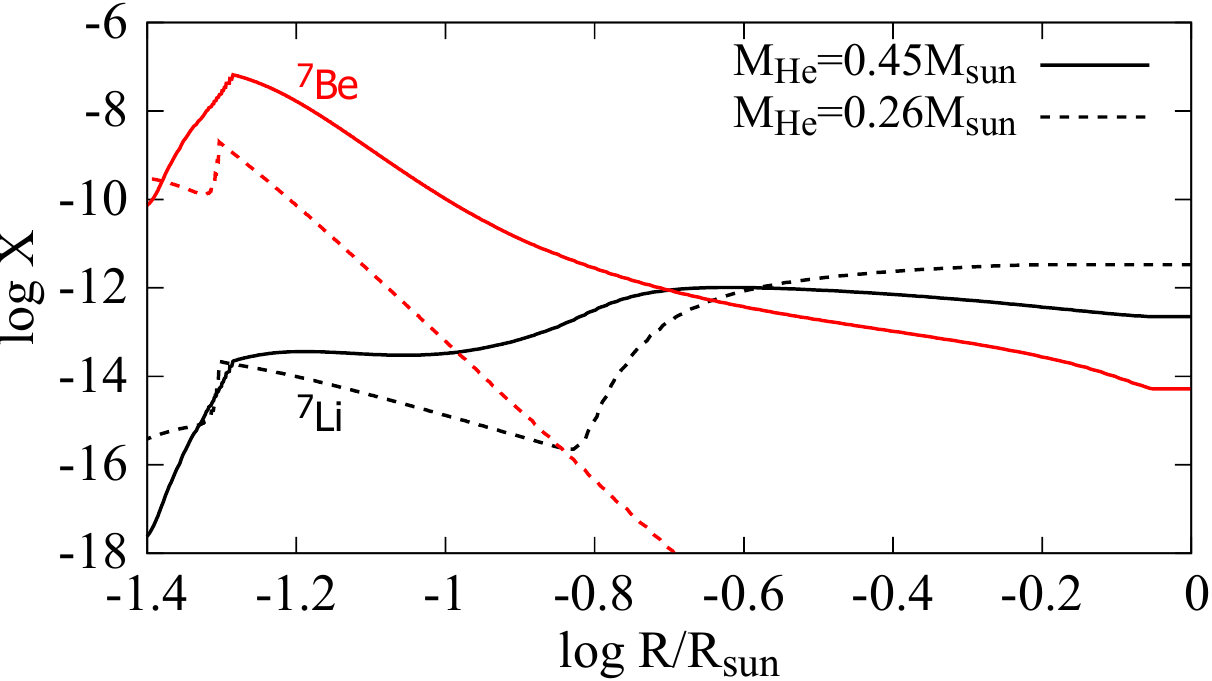}
\caption{The profile of (upper) $D_\mathrm{thm}$, (middle) the mixing and nuclear reaction timescales, and (lower) the abundances of $^7$Be and $^7$Li. The NMM is not adopted, and $\alpha_\mathrm{thm}=100$ is assumed. The solid lines show the model when $M_\mathrm{He}=0.45M_\odot$ and the broken lines show the model when $M_\mathrm{He}=0.26M_\odot$. \label{fiducial}}
\end{figure}

Fig. \ref{fiducial} shows the profile of the $D_\mathrm{thm}$, the mixing and nuclear reaction timescales, and the $^7$Be and $^7$Li abundances in the region where $^7$Be is produced. The solid lines indicate the time when $M_\mathrm{He}=0.45M_\odot$ and the broken lines indicate the time when $M_\mathrm{He}=0.26M_\odot$. {$^7$Be is produced by $^3$He$(\alpha,\;\gamma)^7$Be at $R\sim10^{-1.3}R_\odot$.} It is seen that thermohaline mixing becomes more effective as the core grows. This is because the density in this region decreases as a function of time. The lower density leads to a larger $K$ as we can see from Eq. (\ref{K}) and thus a larger $D_\mathrm{thm}$ \citep{2015MNRAS.446.2673L}.

This efficient mixing below the convective envelope explains why $A$(Li) decreases after reaching the RG branch bump and increases near the TRGB.  In order for the CF mechanism to work, mixing in this region should be efficient enough to convey $^7$Be to the bottom of the convective envelope. When $M_\mathrm{He}=0.26M_\odot$, $^7$Be decays into $^7$Li and is destroyed by the proton capture before being conveyed to the convective envelope because $t_\mathrm{mix}$ is much longer than {$t_\mathrm{dest}$} in $R<10^{-0.2}R_\odot$. Rather, since $^7$Li in the envelope diffuses into the thermohaline region, $A$(Li) decreases in time. On the other hand, when $M=0.45M_\odot$, $t_\mathrm{mix}$ becomes shorter than before. As a result, a part of $^7$Be is conveyed to the envelope before it decays and $A$(Li) increases.

\subsection{Dependence on the NMM}
We perform stellar evolution calculations with $\mu_{12}=1-5$, where $\mu_{12}=\mu_\nu/(10^{-12}\mu_\mathrm{B})$. The adopted parameters and results are summarized in Table 1. $A(\mathrm{Li})$ and $L$  at the TRGB increase when a larger value of $\mu_\nu$ is adopted. The argument that TRGB stars become luminous when $\mu_\nu$ is adopted has been used to constrain $\mu_\nu$ \citep[e.g.][]{1996slfp.book.....R,2015APh....70....1A}. 
\begin{table}
\centering
 \begin{tabular}{cc|ccc}
 $\alpha_\mathrm{thm}$&$\mu_{12}$&$M_\mathrm{He,TRGB}/M_\odot$&$\log (L/L_\odot)_\mathrm{TRGB}$&$A(\mathrm{Li})_\mathrm{RC}$\\\hline\hline
 1000&0&0.467&3.39&$-1.74$\\
1000&1&0.471&3.41&$-1.60$\\
1000&2&0.480&3.46&$-1.26$\\
1000&3&0.490&3.51&$-0.91$\\
1000&4&0.500&3.56&$-0.56$\\
1000&5&0.509&3.61&$-0.28$\\\hline
100&0&0.467&3.39&$-1.08$\\
100&1&0.471&3.41&$-0.98$\\
100&2&0.480&3.46&$-0.75$\\
100&3&0.490&3.51&$-0.46$\\
100&4&0.500&3.56&$-0.18$\\
100&5&0.509&3.61&0.07\\\hline
50&0&0.467&3.39&$-0.73$\\
50&1&0.471&3.41&$-0.70$\\
50&2&0.480&3.46&$-0.57$\\
50&3&0.490&3.51&$-0.36$\\
50&4&0.500&3.56&$-0.13$\\
50&5&0.509&3.61&0.10\\
\end{tabular}
\caption{The parameters of the models. The zero age main sequence mass and the initial metallicity are fixed to $M_\mathrm{ZAMS} =1 M_\odot$ and $Z=0.0148$, respectively. $\alpha_\mathrm{thm}$ is the thermohaline coefficient, $\mu_{12}$ is the NMM, $M_\mathrm{He,TRGB}$ is the mass of the helium core at the TRGB, $L_\mathrm{TRGB}$ is the luminosity at the TRGB, and $A(\mathrm{Li})_\mathrm{RC}$ is the lithium abundance of RC stars. Since $A(\mathrm{Li})$ decreases during the evolution of RC stars, the initial values just after the helium flash are shown.}
\end{table}

The temperature in the core gradually increases as the star evolves because the gravitational energy is released. The energy released near the TRGB is estimated to be $\sim100\;\mathrm{erg\;g^{-1}\;s^{-1}}$ \citep[e.g. Section 2.5.2 in][]{1996slfp.book.....R}. Enhanced energy loss rates in the center 
can result in a more rapid evolution to advanced
stellar burning and growth of the helium core.
If we assume $\mu_{12}=5$, the additional energy loss rate induced by the NMM at the stellar center is $\epsilon^\mu_\mathrm{plas}\approx6\;\mathrm{erg\;g^{-1}\;s^{-1}}$ {just before the onset of the helium-core flash}.  Since this additional energy loss is not negligible compared with the released gravitational energy, the helium flash would be delayed.

\begin{figure}
\centering
\includegraphics[width=8.5cm]{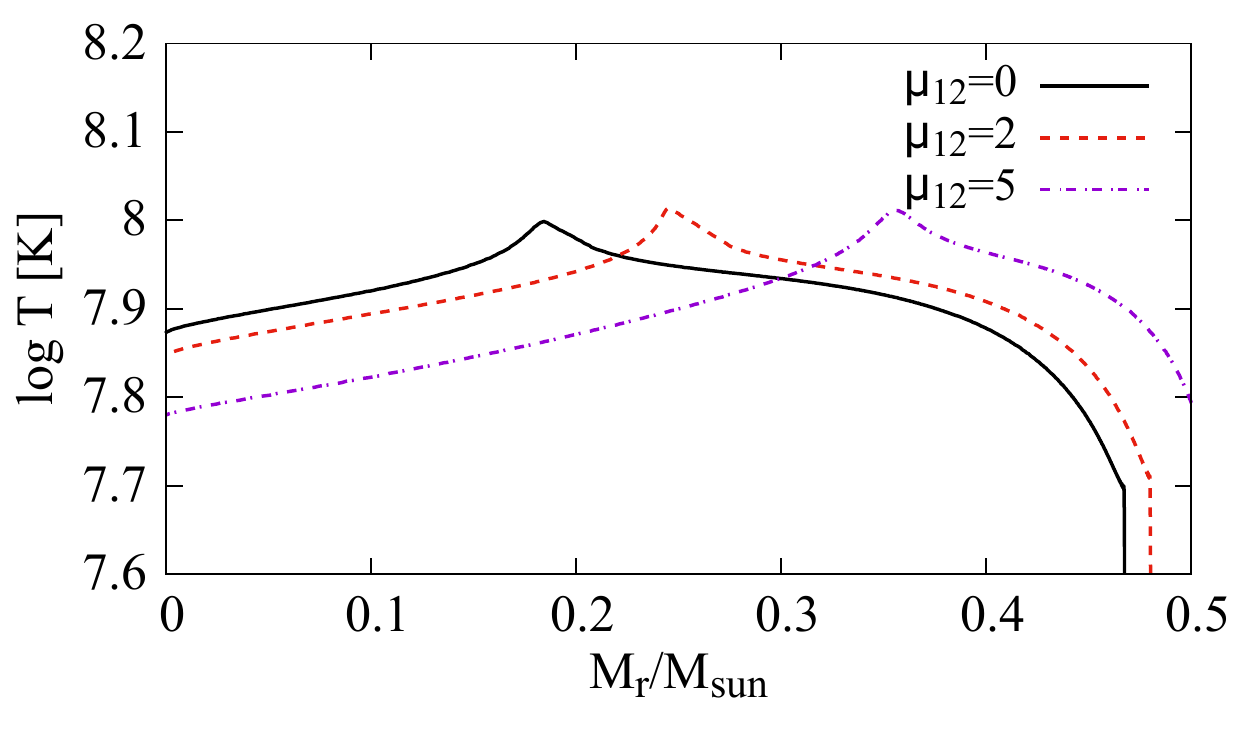}
\caption{The temperature profile in the  core when the helium luminosity reaches $L_\mathrm{He}=100L_\odot$. The solid line shows the model without the NMM and the other lines adopt $\mu_{12}=2$ and 5.  \label{flash}}
\end{figure}

The additional energy loss affects the structure of the helium core just before the helium flash. Fig. \ref{flash} shows the temperature profile in the helium core when the helium luminosity $L_\mathrm{He}$ reaches $100L_\odot$. The solid line shows the fiducial model with $\mu_{12}=0$ and the other lines show the models with $\mu_{12}=2$ and 5. {Because neutrinos carry away energies from the center, the temperature inversion is developed in the core. As a result,  $T$ is slightly higher in an off-center shell as seen in Fig. \ref{flash}.} 
Since the core is degenerate and helium burning is highly sensitive to the temperature, the helium flash is ignited in the hottest shell. When the NMM is adopted, the location of the helium flash shifts outward because the energy loss  is higher \citep{1978ApJS...36..405S,1992A&A...264..536R}.

\begin{figure}
\centering
\includegraphics[width=8.5cm]{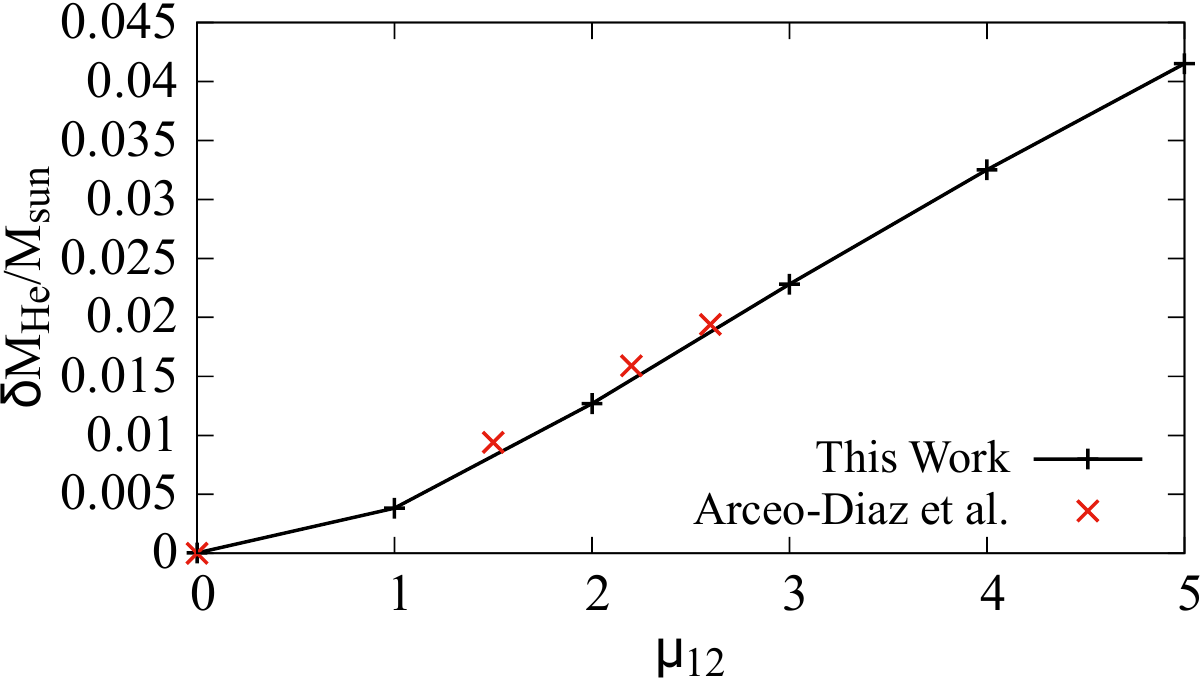}
\caption{The core mass as a function of the NMM. The definition of $\delta M_\mathrm{He}$ is shown in Eq. (\ref{deltam}). The red crosses are $1M_\odot$ models with $Z=0.01$ calculated by \citet{2015APh....70....1A}. \label{coremass1}}
\end{figure}

The previous works \cite[e.g.][]{1996slfp.book.....R} showed that core mass at the TRGB is increased by the additional energy loss. Fig. \ref{coremass1} shows the increase of helium core mass 
\begin{equation}
    \delta M_\mathrm{He}=M_\mathrm{He}(\mu_{12})-M_\mathrm{He}(\mu_{12}=0),\label{deltam}
\end{equation}
where $M_{\rm He}(\mu_{12})$ is the core mass when the helium flash occurs. The solid line shows our model and the red crosses are $1M_\odot$ models with $Z=0.01$ in a previous study \citep{2015APh....70....1A}.  Although physical inputs to the models are different, our result is similar to the previous calculation.


\begin{figure}
\centering
\includegraphics[width=8.5cm]{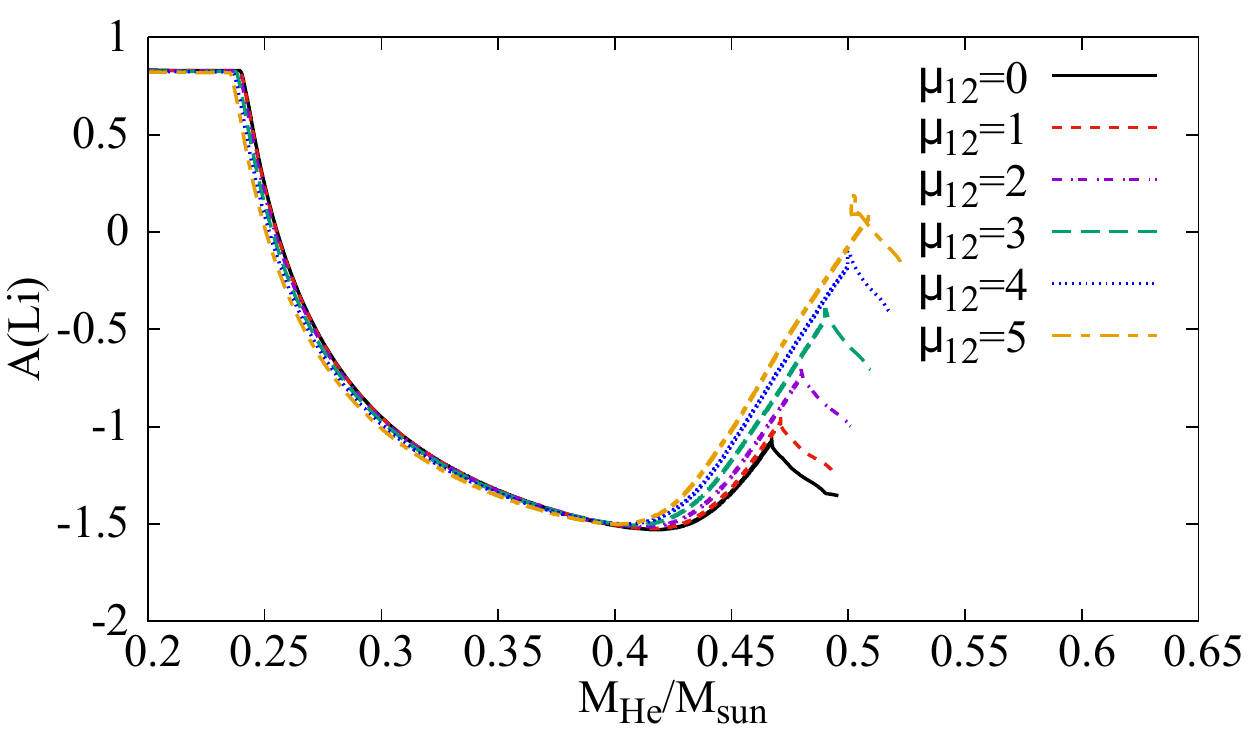}
\caption{The evolution of the surface lithium abundance as a function of the helium core mass. The thermohaline coefficient is fixed to $\alpha_\mathrm{thm}=100$. \label{coremass}}
\end{figure}

Fig. \ref{coremass} shows the evolution of $A(\mathrm{Li})$ as a function of the helium core mass $M_\mathrm{He}$ when $\alpha_\mathrm{thm}=100$. The peaks around $M_\mathrm{He}\sim0.5M_\odot$ correspond to the helium flash. One can confirm that $A(\mathrm{Li})$ at the TRGB is higher when $\mu_\nu$ is larger.

\begin{figure}
\centering
\includegraphics[width=8.5cm]{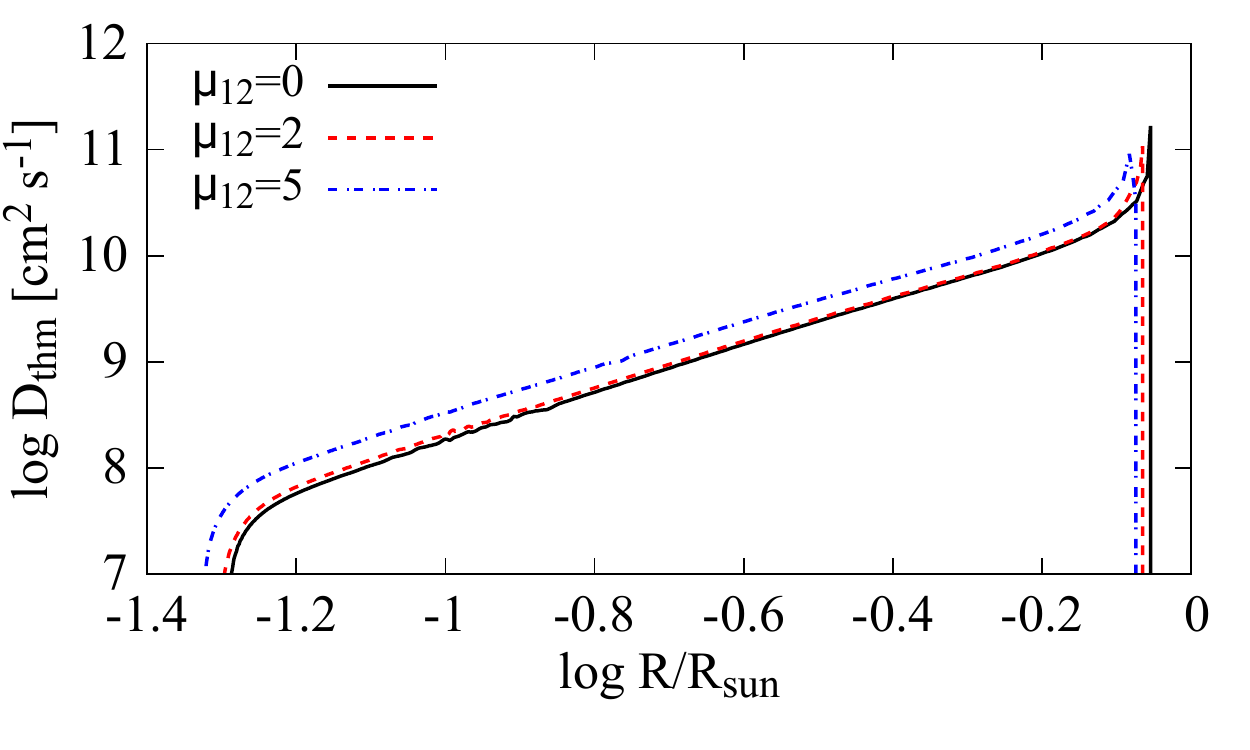}
\includegraphics[width=8.5cm]{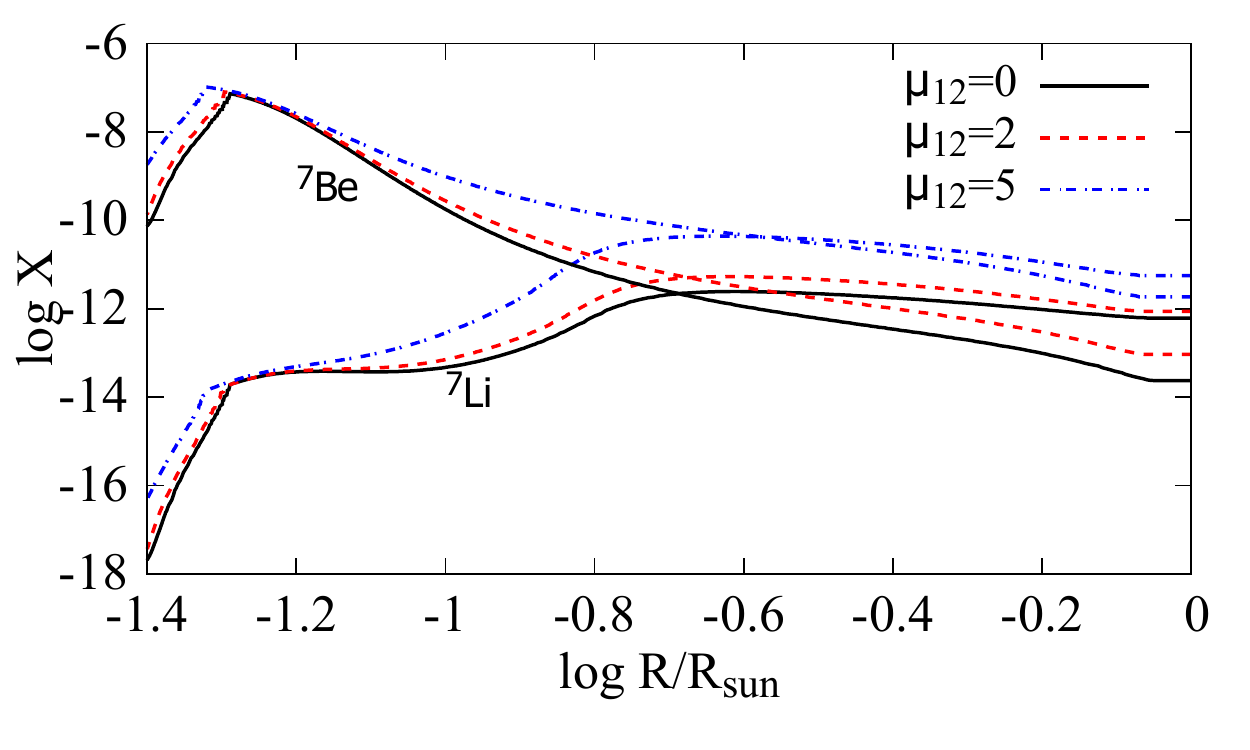}
\caption{Structure of our model at the TRGB with $\mu_{12}=0$, 2, and 5. The upper panel shows the thermohaline diffusion coefficient and the lower panel shows the mass fractions of $^7$Li and $^7$Be as a function of the radius. The thermohaline coefficient is fixed to $\alpha_\mathrm{thm}=100$. }
\label{fig:_TRGB}
\end{figure}

The physical mechanisms of the lithium enhancement are twofold. When $\mu_\nu>0$ is adopted, the helium core on the ignition of the helium flash becomes heavier because of the additional energy loss. Because the flash is delayed, the CF mechanism can continue to produce more $^7$Li and $A(\mathrm{Li})$ in RC stars becomes higher. The more massive core leads to a smaller density above the hydrogen burning shell and hence a larger thermal conductivity \citep{2015MNRAS.446.2673L}. Since $D_\mathrm{thm}$ increases as a fuction of conductivity, thermohaline mixing at the TRGB becomes more efficient with a larger NMM. Fig. \ref{fig:_TRGB} shows the profile of $D_\mathrm{thm}$ and the mass fractions of $^7$Be and $^7$Li at the TRGB. It is seen that $D_\mathrm{thm}$ is larger if the NMM is larger.  Therefore the CF mechanism can convey $^7$Be to the convective envelope more effectively and  $A(\mathrm{Li})$ in RC stars becomes higher.

\begin{figure}
\centering
\includegraphics[width=8.5cm]{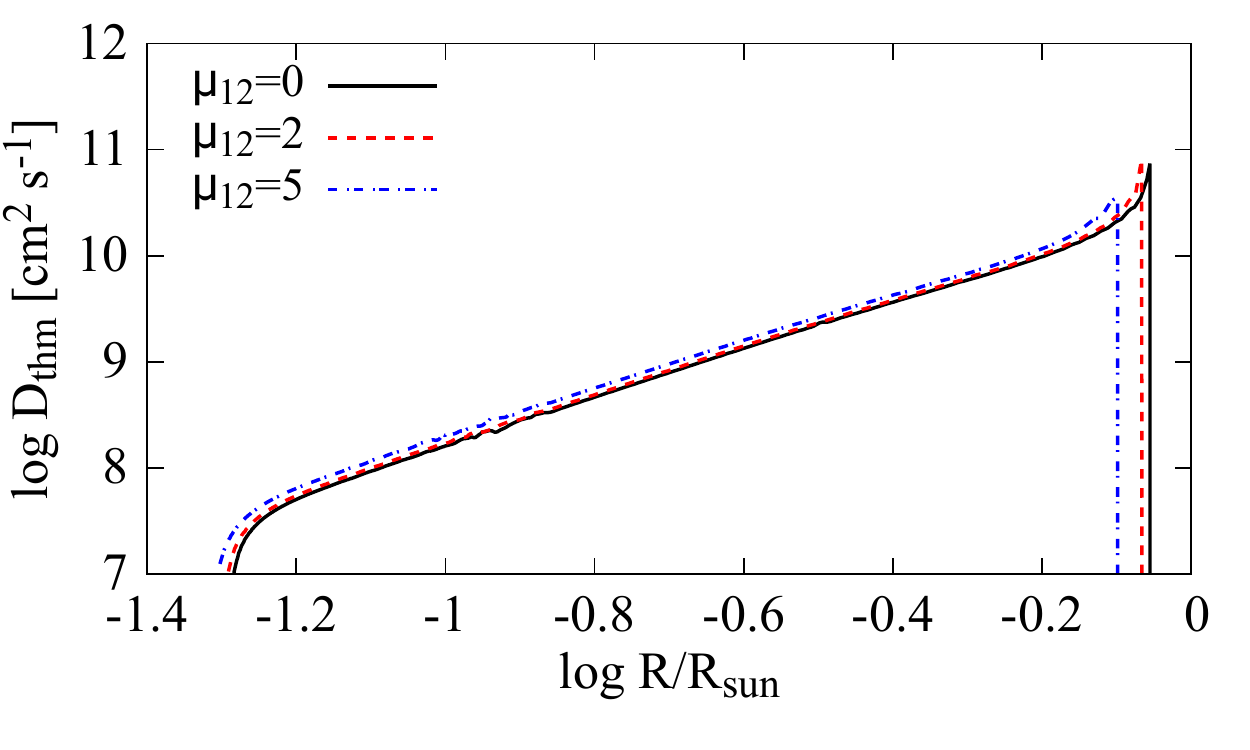}
\includegraphics[width=8.5cm]{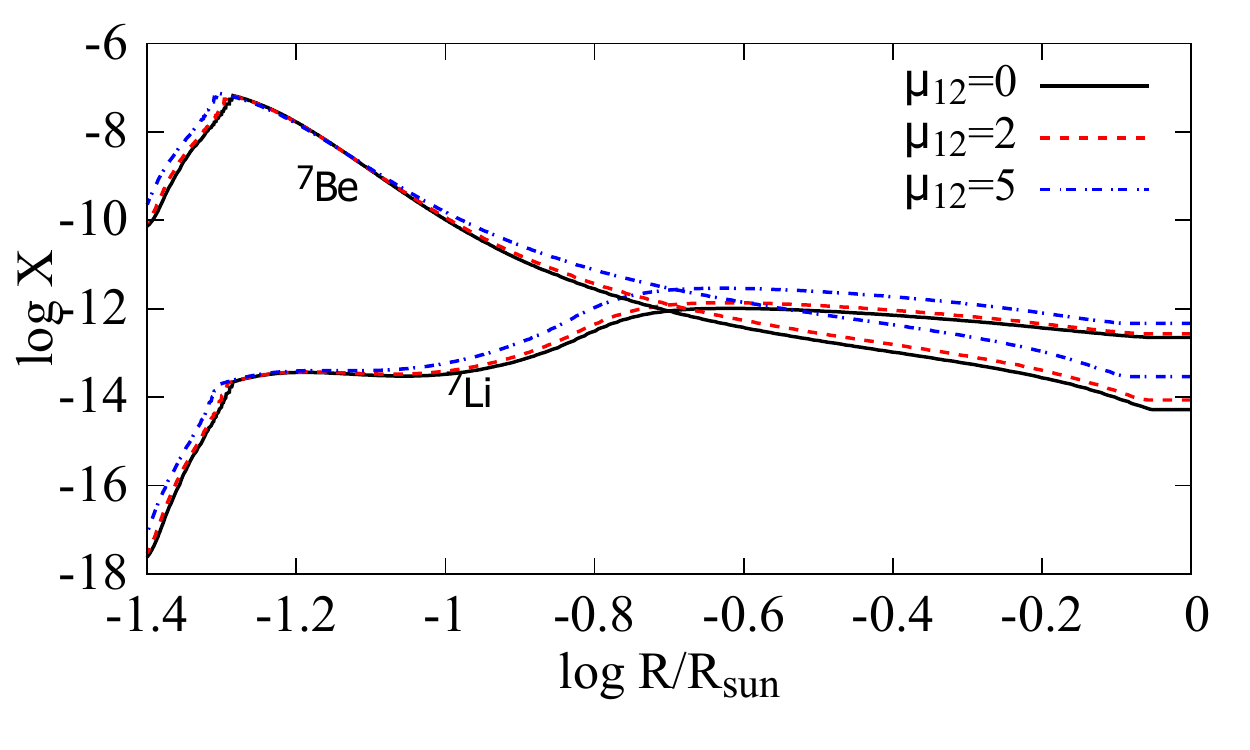}
\caption{Structure of our model when $\mu_{12}=0$, 2, and 5 and  $M_\mathrm{He}=0.45M_\odot$. The upper panel shows the thermohaline diffusion coefficient and the lower panel shows the mass fractions of $^7$Li and $^7$Be as a function of the radius. The thermohaline coefficient is fixed to $\alpha_\mathrm{thm}=100$. \label{structure}}
\end{figure}

It is seen from Fig. \ref{coremass} that $A$(Li) starts to deviate from the standard model even before the helium flash. This is explained by changes of stellar structure induced by the NMM. Fig. \ref{structure} shows the thermohaline diffusion coefficient $D_\mathrm{thm}$ and the mass fractions of $^7$Li and $^7$Be for the models with $\mu_{12}=0$, 2, and 5 in the region where  thermohaline mixing is effective. In this figure, the helium core mass is fixed to $M_\mathrm{He}=0.45M_\odot$. When a larger NMM is adopted, the radius of the helium core becomes smaller and the density in the envelope decreases. 
Since $^7$Be produced via $^3$He($\alpha$,$\gamma$) is conveyed to the outer region by thermohaline mixing, the more efficient mixing leads to a larger $A$(Li).

Recently, the GALAH (Galactic Archaeology with HERMES) survey second data release (DR2) provided spectroscopic data of 342,682  stars in the Milky Way \citep{2018MNRAS.478.4513B}. \citet{2020NatAs.tmp..139K} selected stars with $\log L/L_\odot\in[1.55,\;1.85]$ and the effective temperature $T_\mathrm{eff}\in[4650,\;4900]$ K from the GALAH DR2 samples and identified them as RC stars. \citet{2020NatAs.tmp..139K} used GALAH samples that overlap with an astroseismic catalog \citep{2018ApJ...858L...7T} to distinguish RC and RG stars. They concluded that the contamination of RGs in their RC samples accounts for only $\sim10\%$. 

\citet{2020NatAs.tmp..139K} found that the lithium abundance in RC stars is distributed around $A(\mathrm{Li})\sim0.71\pm0.39$. This ubiquitous enhancement of lithium has not been predicted by stellar models. When the NMM is not adopted in our model, the lithium abundance in RC stars is only $A(\mathrm{Li})=-1.08\;(-0.73)$ when $\alpha_\mathrm{thm}=100$ (50). We find that, if $\mu_{12}=5$ is adopted, $A(\mathrm{Li})$ reaches 0.07 (0.10), which is closer to the observed $A$(Li). When $\alpha_\mathrm{thm}$ is larger, $^7$Li is destroyed to a greater extent after the RG branch bump and thus $A$(Li) in RC stars becomes smaller. Although $A$(Li) is not sufficiently large when $\mu_{12}=2-4$ in both cases, the discrepancy in $A$(Li) becomes smaller if the NMM is adopted. The additional energy loss induced by the NMM is thus a candidate of a ubiquitous mechanism of the high $A$(Li) in RC stars.

Traditionally, giants with $A(\mathrm{Li})>1.5$ have been called lithium-rich giants \citep{1989ApJS...71..293B}. It is difficult to explain such extremely high lithium abundances with the NMM only. \citet{2020NatAs.tmp..139K} point out that lithium-rich giants with $A(\mathrm{Li})>1.5$ account only for $\sim3.0\%$ of RC stars. The rare population implies another mechanism which works only in a certain kind of stars.
\subsection{Dependence on Stellar Mass}
\begin{figure}
\centering
\includegraphics[width=8.5cm]{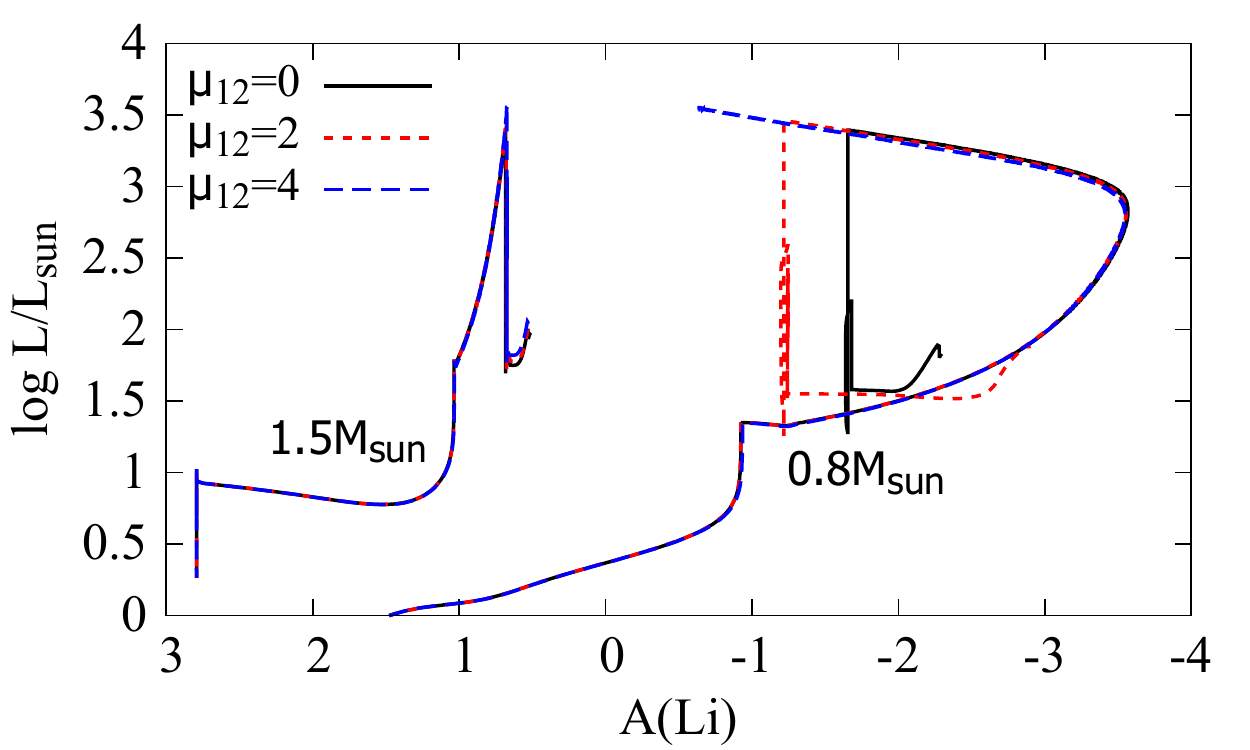}
\caption{The evolution of our $1.5M_\odot$ and $0.8M_\odot$ models with $\mu_{12}=0$, 2, and 4 and $Z=0.148$ in the $L-A$(Li) plane. Efficiency of thermohaline mixing is fixed to $\alpha_\mathrm{thm}=50$.  \label{mass}}
\end{figure}

In Section \ref{result1}, we investigated $1M_\odot$ models in detail because the stellar samples \citep{2020NatAs.tmp..139K} are distributed around $(1.0\pm0.4)M_\odot$. In this section, we study $1.5M_\odot$ and $0.8M_\odot$ models to see dependence on stellar masses. Fig. \ref{mass} shows the evolutionary tracks in the plane of the luminosity $L$ and $A$(Li) in the models with $\mu_{12}=0$, 2, and 4. Efficiency of thermohaline mixing is fixed to $\alpha_\mathrm{thm}=50$. In the $1.5M_\odot$ models, the lithium enhancement near the TRGB is not seen. Also, the effect of the NMM is smaller than in the $1M_\odot$ case. {This is because the lithium depletion due to the first dredge-up is less significant in heavier stars. As a result, $A$(Li) in a more massive RG is higher. Although thermohaline mixing works even in the $1.5M_\odot$ model, the relative change in $A$(Li) during the evolution along the RG branch is smaller and the contribution of the $^7$Be transport to $A$(Li) is negligible. Such larger values of $A$(Li) in heavier RGs are reported also in previous works \citep[e.g.][]{2020arXiv200901248S}.}

On the other hand, the $0.8M_\odot$ models show large enhancements of $A$(Li) even if $\mu_{12}=0$. The lithium enhancement becomes larger when the NMM is considered as we saw in the $1M_\odot$ models. The model with $\mu_{12}=4$ does not experience the helium flash. This is  because the larger luminosity near the TRGB results in a higher mass loss rate and thus the model loses the hydrogen envelope before the helium flash. {Such a star would become a massive helium white dwarf. Although the lack of the flash would contradict observations of globular clusters, it depends on mass loss rates, which are still uncertain. We performed additional calculations of $0.8M_\odot$ models with $(\eta,\;Z)=(0.3,\;0.007)$, (0.2, 0.0148), and (0.2, 0.007) to check the dependence on the mass loss. We found that all of the additional models undergo the flash because of smaller mass loss rates. Therefore non-existence of single massive helium white dwarfs does not necessarily constrain the NMM.}
\subsection{Dependence on Metallicity}
\begin{figure}
\centering
\includegraphics[width=8.5cm]{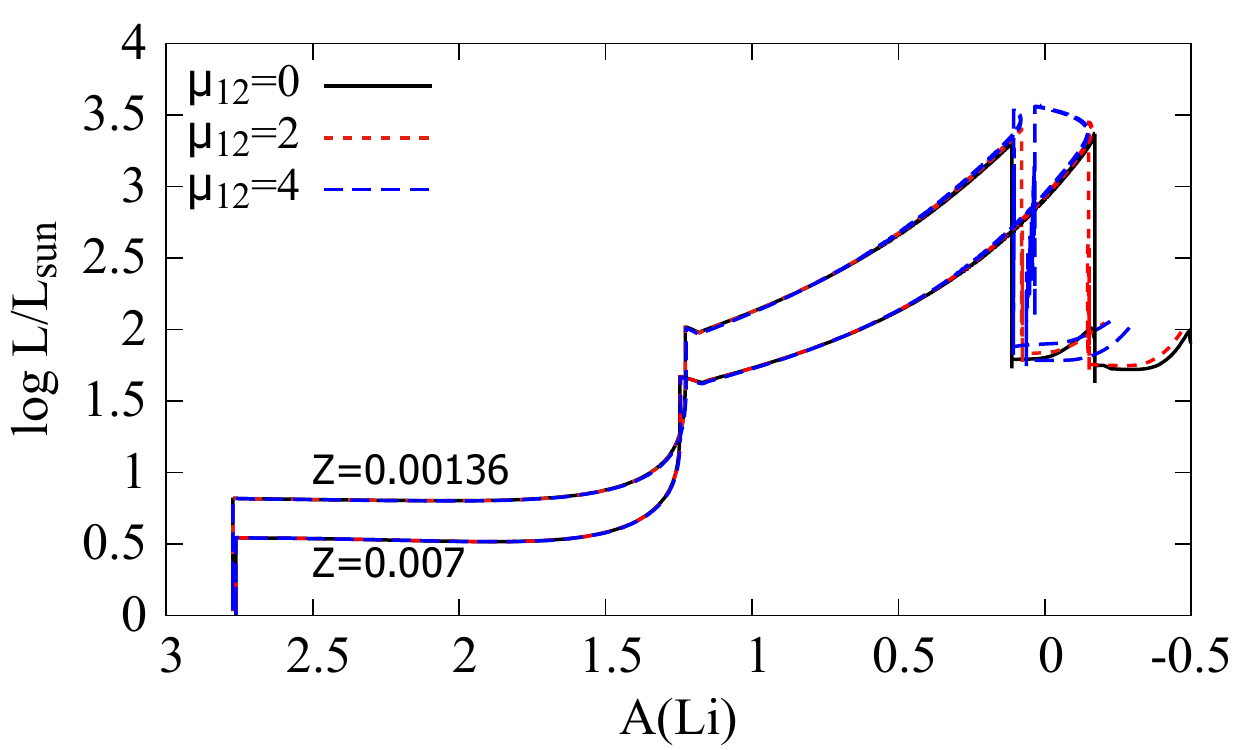}
\caption{The evolution of our $1M_\odot$ models with $\mu_{12}=0$, 2, and 4 and $Z=0.00136$ and 0.007 in the $L-A$(Li) plane. Efficiency of thermohaline mixing is fixed to $\alpha_\mathrm{thm}=50$.  \label{metal}}
\end{figure}

In Section \ref{result1}, we investigated Solar metallicity models in detail because the stellar samples are distributed around $\mathrm{[Fe/H]}=(-0.1\pm0.2)$. In this section, we study $Z=0.00136$ and 0.007 models to see dependence on metallicities. The metallicity $Z=0.00136$ corresponds to the metal-poor environment in the globular cluster M5 \citep{2013A&A...558A..12V} and $Z=0.007$ is between the Solar and M5 metallicities. Fig. \ref{metal} shows the evolutionary tracks in the plane of $L$ and $A$(Li) in $Z=0.00136$ and 0.007 models. Efficiency of thermohaline mixing is fixed to $\alpha_\mathrm{thm}=50$. It is seen that, if $\mu_{12}$ and $\alpha_\mathrm{thm}$ are fixed, $A$(Li) in RC stars is larger when $Z$ is lower. In the metal-poor models, the lithium enhancement near the TRGB is not seen if $\mu_{12}\leq2$. In the $Z=0.00136$ models, $A$(Li) is not enhanced even if $\mu_{12}=4$. In the $Z=0.007$ model, $A$(Li) slightly increases near the TRGB when $\mu_{12}=4$, although the effect of the NMM is smaller than in the $Z=Z_\odot$ case.
\section{Discussion}
In this paper, we discussed the effects of additional energy loss channels on $A$(Li) in RC stars. In the special case of NMM we found that the production of $^7$Li near the TRGB is activated when $\mu_{12}=1-5$ is adopted.  Although this value of the neutrino magnetic moment is smaller than current limits obtained from the reactor experiments, it exceeds the recent astrophysical upper limits. It is therefore difficult to solve the lithium problem only with the NMM. However, the additional energy loss can be induced by other physics like extra dimensions \citep{2000PhLB..481..323C} and axion-like particles \citep{1987PhRvD..36.2211R,2014PhRvL.113s1302A}. Since they are expected to result in the similar enhancement of $A$(Li), they can be a candidate of the mechanism of the lithium enhancement.

The destruction and production of $^7$Li are dependent on deep mixing including thermohaline mixing \citep{2007A&A...467L..15C,2015MNRAS.446.2673L} and magnetic buoyancy \citep{2007ApJ...671..802B}. It is desirable to investigate these mechanisms in detail.

The enhancement of energy loss rate from neutrino emission affects Li abundances on stellar surfaces through a change in stellar structure, including He core mass  and the mixing time scale. This characterizes the current theoretical prediction distinguished from other possibilities. For example, in addition to the $^7$Be production via the $^3$He($\alpha$,$\gamma$) reaction operating deep inside the stars, stellar surfaces can be polluted from outside by accretion of companion stellar ejecta or nucleosynthesis via flare-accelerated nuclei on stellar surfaces. If the observed high abundances of Li originate from nucleosynthesis in companion asymptotic giant branch stars \citep{2010MNRAS.402L..72V}, observed stars can have enhanced abundances of carbon and $s$-nuclei. On the other hand, if nuclear reactions of flare-accelerated nuclei \citep{2007A&A...469..265T} are providing $^{6,7}$Li, the isotopic fraction of $^6$Li is expected to be high. Furthermore, the observed Li-rich stars must be associated with very strong flare activities and simultaneous production of Be and B. In this way, respective possibilities are associated with different astronomical observables to be measured in future.

\begin{figure}
\centering
\includegraphics[width=8.5cm]{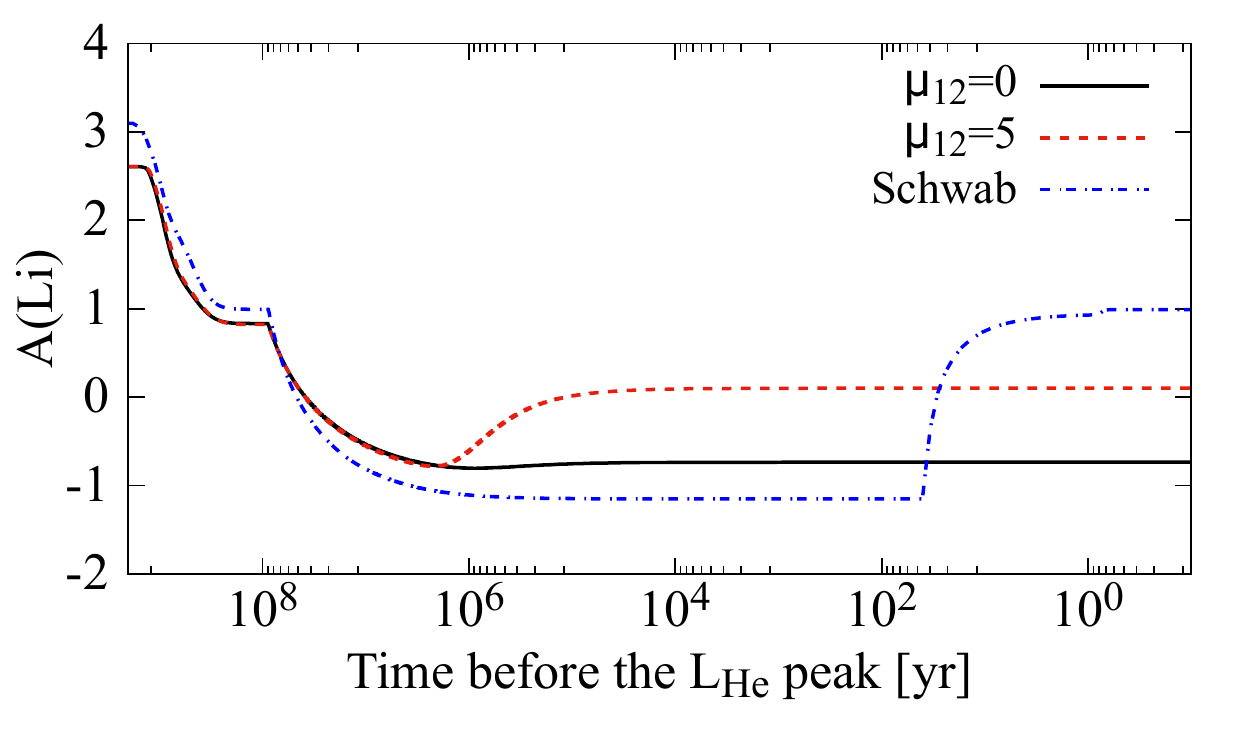}
\caption{The evolution of $A$(Li) as a function of time. The black and red lines show $A$(Li) in our model with $\mu_{12}=0$ and 5, respectively. The blue line shows the $1M_\odot$ model with helium-flash induced mixing \citep{2020arXiv200901248S}. Efficiency of thermohaline mixing is fixed to $\alpha_\mathrm{thm}=50$ in our model.  \label{timescale_comparison}}
\end{figure}
A recent study \citep{2020arXiv200901248S} showed that the lithium problem in RC stars can be solved if mixing is induced by the helium flash. He assumed that mixing with the diffusion coefficient $D=10^{12}\;\mathrm{cm^2\;s^{-1}}$ is induced when the helium luminosity $L_\mathrm{He}$ exceeds $10^4L_\odot$. A major difference between  his and our models is the timescale of the surface lithium enhancement. Fig. \ref{timescale_comparison} shows $A$(Li) as a function of time. The black and red lines show our model while the blue line shows the $1M_\odot$ model in \citet{2020arXiv200901248S}. It is seen that $A$(Li) is enhanced in $\sim100$ yr during the helium flash in 
his model, while it occurs $\sim1$ Myr before the flash in our model. Although the time evolution of $A$(Li) near the TRGB is dependent on model parameters including $\alpha_\mathrm{thm}$, it is robust that the lithium enhancement in our model is much slower than the model in  \citet{2020arXiv200901248S}. Hence it might be possible to distinguish the two models by searching for lithium-rich RGs with $A(\mathrm{Li})\sim 0$ near the TRGB. The model in \citet{2020arXiv200901248S} predicts that it would be impossible to find such stars, while our model predicts that such stars would be found if $A$(Li) values of a sufficient number of stars are observed.
\section*{Acknowledgements}
We thank Yerra Bharat Kumar and Gang Zhao for providing observational data. K.M. is supported 
by JSPS 
KAKENHI Grant Number JP19J12892. 
T.K. is supported in part by Grants-in-Aid for Scientific Research of JSPS (17K05459, 20K03958).  A.B.B. is supported in part by the U.S. National Science Foundation Grants No. PHY-1806368 and 
PHYS-2020275.  M.A.F. is supported by National Science Foundation Grant No. PHY-1712832 and by NASA Grant No. 80NSSC20K0498.  M.A.F. and A.B.B. acknowledge support from the NAOJ Visiting Professor program.

\section*{Data Availability}

The code and the inlist used in this paper are available at https://doi.org/10.5281/zenodo.4281702.








\bsp	
\label{lastpage}
\end{document}